\documentclass [prl,aps,twocolumn,superscriptaddress,floatfix,amsmath,amssymb] {revtex4-2}
\usepackage{graphicx}
\usepackage{hyperref}
\usepackage[caption=false]{subfig}
\usepackage[capitalize]{cleveref}
\usepackage{mathtools}
\DeclarePairedDelimiter\abs{\lvert}{\rvert}

\begin{document}
\title{Unconventional discontinuous transitions in isospin systems}
\author{Zachary M. Raines}
\affiliation{School of Physics and Astronomy and William I. Fine Theoretical Physics Institute, University of Minnesota, Minneapolis, MN 55455, USA}
\author{Leonid I. Glazman}
\affiliation{Department of Physics, Yale University, New Haven, CT 06520, USA}
\author{Andrey V. Chubukov}
\affiliation{School of Physics and Astronomy and William I. Fine Theoretical Physics Institute, University of Minnesota, Minneapolis, MN 55455, USA}
\begin{abstract}
We show that two-dimensional
fermions with
dispersion $k^2$ or $k^4$
undergo
a first-order
Stoner transition to a fully spin-polarized state
despite that the spin susceptibility diverges at the critical point.
We extend our analysis to systems with dispersion $k^{2\alpha}$ and spin and valley isospin and show that
there is a cascade of instabilities into fractional-metal states with some
electron bands fully depleted;
narrow intermediate ranges of partially-depleted bands
exist for $\alpha<1$ or $\alpha > 2$.
The susceptibility becomes large near each transition.
We discuss applications to biased
bilayer and tri-layer graphene
and moir\'e systems.
\end{abstract}

\maketitle

\emph{Introduction.}~Quasi-two-dimensional materials with spin and valley isospin degrees of freedom are currently attracting a lot of attention triggered by
experimental discoveries of qualitatively new physics~\cite{Bistritzer2010,Cao2018,Cao2018a,Andrei2020,Xie2019,Wong2020,Zondiner2020}.
The experiments that chiefly motivated our study are on non-twisted Bernal-stacked bilayer graphene (BBG) and rhombohedral tri-layer graphene (RTG) in a displacement field.
Quantum oscillations and inverse compressibility measurements on BBG and RTG indicated~\cite{Seiler2022,Zhou2022a,Zhou2021,DeLaBarrera2022,Seiler2023,Arp2023,Holleis2023,Zhang2023a} that there is a cascade of spin/valley ordered states and revealed two features, which are seemingly incompatible with each other.
On one hand, there are indications of soft bosonic excitations on the ``paramagnetic'' side of the transition, which normally is an indication of a continuous transition; on the other, in the ordered state some fermionic bands get fully depleted (half-metal and quarter-metal states, in terminology of~\cite{Zhou2021}), which points to strong first-order transitions into fully polarized states.
A first-order transitions into spin/valley
polarized states, accompanied by
the growth of the corresponding susceptibilities,
have been also reported in experiments on a 2D electron gas
AlAs~\cite{Hossain2021,*Hossain2022,*Hossain2020} and detected in variational Monte-Carlo calculations~\cite{Valenti2023}.

These experiments call for better understanding of a Stoner transition in 2D.
The Stoner model~\cite{Stoner1939} has been used for decades to describe itinerant ferromagnetism.
In its simplest form, it describes parabolicly dispersing electrons with a contact repulsive interaction $U$.
Mean field theory shows that in three-dimensions (3D) the system undergoes a second-order transition into a ferromagnet at a critical coupling given by the \emph{Stoner criterion} $\nu_{F}U=1$.
Beyond this transition point, the spin-rotation symmetry is spontaneously broken, giving rise to a net magnetization and splitting the dispersions for spin-up and spin-down fermions.
Fluctuation effects introduce non-analytic behavior possibly leading to a first-order transition~\cite{Hertz1976,*Millis1993,*Moriya2012,Vojta1999,Chubukov2004,*Efremov2008,*Chubukov2009}, but this first-order transition is a weak one and does not strongly modify band fillings compared to mean-field.

\begin{figure}
    \centering
    \includegraphics[width=\linewidth]{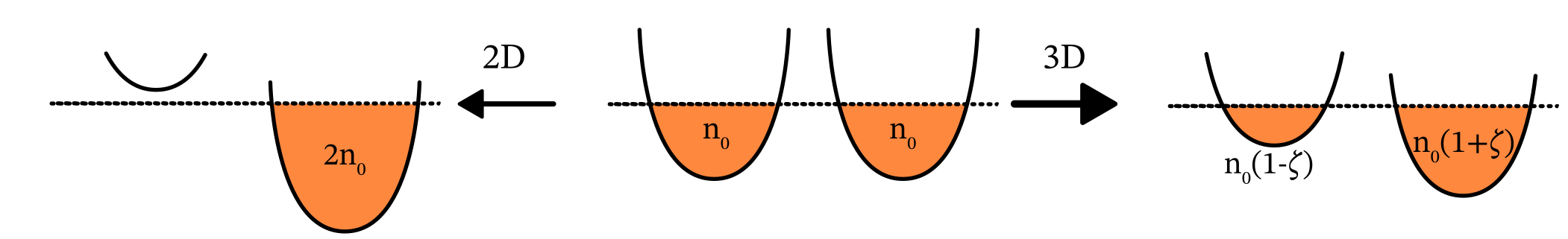}
    \caption{(Color online)
        Filling of the parabolic spin bands below Stoner instability in 3D and 2D.
        \label{fig:schematic-fillings}}
\end{figure}

In this letter, we show that in 2D the Stoner transition is different already at the mean-field level.
Namely, for a parabolic dispersion, the transition is first order to a half-metal state with
full polarization and a fully depleted band for one spin polarization,
\cref{fig:schematic-fillings}.
However, the transition occurs at
exactly the critical coupling for the Stoner instability.
This leads to a situation where the magnetic susceptibility diverges as the transition is approached from a paramagnetic state, as would normally be expected for a second-order transition, yet immediately
beyond
the transition the polarization jumps to the maximum possible value.
We argue that this is \emph{not} just the consequence of a density-independent $\nu_F$ for a parabolic dispersion in 2D:\@ to high numerical accuracy the Stoner transition remains the same for a set of models with isotropic dispersion $k^{2\alpha}$ with $1 \leq \alpha \leq 2$.
Moreover, for $\alpha =2$ ($k^4$ dispersion) the behavior is exactly the same as for a parabolic dispersion.
We extend the analysis to systems with spin and valley degrees of freedom with four initially degenerate bands and approximate $\mathrm{SU(4)}$ symmetry.
We derive the full Landau function for an $\mathrm{SU(4)}$ order parameter without expanding in its powers
and show that for anisotropic $\mathrm{SU(4)}$ system, there are two Stoner-type instabilities first into a half-metal state and subsequently into a quarter-metal state with, accordingly, two and three bands out of the total four fully depleted.
Near each instability the corresponding susceptibility diverges and immediately beyond the transition an order parameter jumps to full polarization.
For a system with exact $\mathrm{SU(4)}$ spin-valley symmetry, we show that there is a cascade of transitions into a three-quarter-metal state, a half-metal state, and a quarter metal state.
For $1 \leq \alpha \leq 2$, there are no other states besides these three; for $\alpha <1$ and $\alpha >2$ there are intermediate states with partial polarization, but they exist in very narrow ranges of the coupling.
We apply the results to Bernal bilayer graphene in a displacement field and discuss potential application to moir\'e systems.

\emph{Ferromagnetic Stoner transition in 2D.}~Consider a system of 2D spinfull fermions with parabolic dispersion $k^2/(2m)$ and momentum independent interaction $U$:
\begin{equation}
    \hat{H} =  \sum_{\mathbf{k,\sigma}}c^{\dagger}_{\mathbf{k}\sigma}(\epsilon_{|\mathbf{k}|} - \mu)c_{\mathbf{k}\sigma}
    + U \int d\mathbf{x}\
    c^{\dagger}_{\mathbf{x}\uparrow}
    c^{\dagger}_{\mathbf{x}\downarrow}
    c_{\mathbf{x}\downarrow}
    c_{\mathbf{x}\uparrow}.
\end{equation}
where $\sigma= \uparrow,\downarrow$.
We parameterize the band densities $n_\sigma \equiv n_0 (1 + \sigma \zeta)$ with the polarization $-1\leq \zeta \leq 1$.
The internal energy is simply the sum of the kinetic
$\sum_{\sigma} n_{\sigma}^{2}/2\nu$,
and potential $\frac{U}{4}\left(
    \sum_{\sigma} \sigma n_{\sigma}\right)^{2}$
terms, where $\nu = m/2\pi$ is the density of states per spin.
Since both are quadratic in density, the dimensionless Landau internal energy $F(\zeta)$ (the normalized energy difference between the normal and ordered states) is \emph{exactly} quadratic in the polarization $\zeta$:
$F(\zeta) = \frac{1}{2}\left(1 - \nu U\right)\zeta^{2}$.
~\footnote{
    This is in contrast with the 3D case where for a parabolic dispersion $u_{K} \sim n^{5/3}$.
}
A critical point, where the normal state becomes unstable, is then set by the usual Stoner criterion $\nu U= 1$,
where the magnetic susceptibility $\chi \propto 1/(1- \nu U)$ diverges.
At the same time, immediately upon crossing this point, the system's energy is minimized by setting the order to the largest possible value $\abs{\zeta} =1$, since the energy profile is simply an inverted parabola.
For this $\zeta$, $n_\uparrow =0$ or $n_{\downarrow} =0$, hence one of the two bands gets fully depleted and the system becomes a half-metal, \cref{fig:schematic-fillings}.

\begin{figure}
    \centering
    \includegraphics[width=\linewidth]{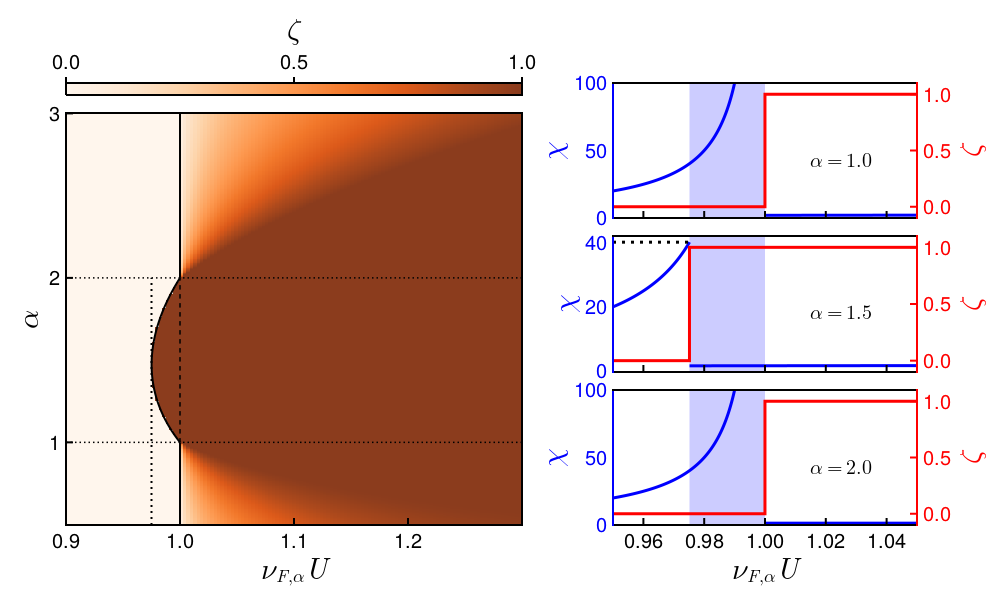}
    \caption{(Color online) Left panel: Phase diagram as a function $\alpha$, the power of the electron dispersion $\epsilon\sim k^{2\alpha}$ and the dimensionless coupling strength $\nu_{F,\alpha}U$.
        Right panel: The static magnetic susceptibility $\chi$ (blue solid line, left y-axis) and polarization $\zeta$ (orange dashed line, right y-axis) as a function of the dimensionless coupling $\nu_{F, \alpha}U$ for a selection of $\alpha$s.
        \label{fig:su2_phase}
    }
\end{figure}

We next extend the analysis to dispersions $\epsilon(|\mathbf{k}|) = c(k^{d}/4\pi)^{\alpha}$ with $\alpha > 0$
(parabolic dispersion is $\alpha =1$).
The respective internal energy $F_\alpha(\zeta)$ has the form
\begin{equation}
    F_\alpha(\zeta)
    =
    \frac{1}{2}
    \left(
    h(\zeta)
    -
    \nu_{F,\alpha} U
    \right)\zeta^{2}
    \label{eq:delta-u}
\end{equation}
where
$\nu_{F,\alpha} = 1/c \alpha n^{\alpha-1}_{0}$,
and
\begin{equation}
    h(\zeta)  \equiv
    \frac{1}{\alpha( \alpha+1)}\frac{
        (1+\zeta)^{\alpha+1}
        + (1-\zeta)^{\alpha+1}
        -2
    }{\zeta^{2}}.
    \label{eq:h}
\end{equation}
The function
$h(\zeta)$ is a constant not only for $\alpha =1$, but also for $\alpha =2$.
In both cases, the Stoner transition then occurs when the susceptibility diverges, and the state immediately beyond the transition is a fully polarized half-metal.
We show the results for other $\alpha$ in \cref{fig:su2_phase}.
Both $\alpha =1$ and $\alpha =2$ separate regions of first-order and continuous transition.
For $1\leq\alpha\leq2$, the transition is first-order to full polarization,
and occurs at
$\nu_{F,\alpha} U
    h(1)=
    2\left(2^{\alpha}-1\right)/\alpha(\alpha+1)$.
While this value is smaller than one, i.e., a first order transition occurs
before the normal state becomes locally unstable, the threshold value remains very close to one: the deviation $1 - h(1)$ does not exceed $0.025$ (see \cref{fig:su2_phase}).
This implies that for all practical purposes it looks like the susceptibility diverges at the onset of a first-order transition, like at $\alpha =1$ and $\alpha =2$.
For $\alpha < 1$ or $\alpha>2$, the system undergoes a second-order phase transition at $ \nu_{F,\alpha}U = 1$.
Still,
the largest value $|\zeta|=1$ is reached at $U$ only slightly larger than $1/\nu_{F,\alpha}$ ($U =1.4/\nu_{F,\alpha}$ for $\alpha =0.5$ and $\alpha =3$)~
\footnote{A strong first order Stoner transition into a half-metal, accompanied by the near-divergence of the susceptibility holds also in 3D, but in a different range of dispersions $k^{2\alpha}$:  between $\alpha =3/2$ and $\alpha =3$ (i.e., for dispersions between $k^3$ and $k^6$).
    For $\alpha =3/2$ and $\alpha =3$, the behavior is the same as for $\alpha =1$ and $\alpha =2$ in a 2D system.}.

\emph{Collective excitations.}~The highly unconventional nature of the Stoner transition in 2D can be also detected by analyzing the dispersion of collective modes.
Consider a parabolic dispersion as an example.
In the paramagnetic state near the instability, the collective excitations in the spin channel are three-fold degenerate Landau overdamped zero-sound modes $\omega \approx -i \delta v_F q$, where $\delta = 1 - \nu_F U$.
In the ordered state, these modes generally split into a transverse Goldstone mode (poles in the dynamical susceptibility at $\omega = \pm q^2/(2M) -i0^{+}$) and a longitudinal mode.
In a 3D system, where $\zeta^2 \propto |\delta|$, the longitudinal mode bounces back into the lower half-plane
($\omega = -2i |\delta|v_F q$) indicating that the ordered state with continuously evolving magnetization and band filings is stable.
In 2D, however, for any partial polarization $|\zeta| <1$, the longitudinal pole continues into the upper half-plane indicating that such a state is unstable.
The system becomes stable only at $|\zeta|=1$, when one of the bands gets depleted and the longitudinal mode vanishes (the range of $q$ where it exists shrinks to zero).

\emph{Spin-valley systems.}~We consider the case, relevant to BBG, RTG, transition metal dichalcogenides, and electron gases in a 2D quantum well, when there are two valleys centered at symmetry related $K$ and $K'$ points in the Brillouin zone, well separated in momentum space.
The low energy theory can be written
as a fully anti-symmetrized action to explicitly treat all channels on the same footing
\begin{multline}
    S =  \sum_{k}
    \bar{\psi}_{k}(i\epsilon_{n} - \epsilon_{\abs{\mathbf{k}}} + \mu)\hat{\tau}_{0}\hat{\sigma}_{0}\psi_{k}\\
    - \frac{1}{8}T^{2}\sum_{kk'q} \sum^{15}_{\gamma=1} U_{\gamma}\bar{\psi}_{k+\frac{q}{2}}\hat{\Gamma}_{\gamma}\psi_{k-\frac{q}{2}}\bar{\psi}_{k'-\frac{q}{2}}\hat{\Gamma}_{\gamma}\psi_{k'+\frac{q}{2}},
    \label{eq:Smodel-su4}
\end{multline}
where the summations and fermion operator indices implicitly include the spin ($\sigma$) and valley ($\tau$) variables,
$\hat{\Gamma}_{\gamma}$ are $4\times4$ matrices generating $\mathrm{SU(4)}$,
and $U_\gamma$ are the corresponding interactions.
The
Coulomb interaction in the total density channel plays no role in Stoner instabilities and we do not display it in \cref{eq:Smodel-su4}.
The 15 generators of $\mathrm{SU(4)}$ are one-component valley polarization, $2$ component
inter-valley coherence order $\hat{\sigma}_{0}\hat{\tau}_{x,y}$ (a charge-density-wave with momentum $K-K'$), $3+3=6$ component valley-symmetric $\hat{\sigma}_{i}\hat{\tau}_{0}$ and valley-staggered ferromagnetism $\hat{\sigma}_{i}\hat{\tau}_{z}$, and $6$-component inter-valley spin coherence order $\hat{\sigma}_{i}\hat{\tau}_{x,y}$ (a spin density wave).
Instability conditions for each of these states have been investigated theoretically by several groups~\cite{ghazaryan2021unconventional,chatterjee2022inter,Chichinadze2022,Chichinadze2022a,You2022,Szabo2022,Xie2023,
    Dong2023,Dong2023b,Blinov2023,*Blinov2023a}.
Exact $\mathrm{SU(4)}$ symmetry develops when all $U_\gamma$ are equal to $U$, which is the case when inter-valley exchange interaction with momentum transfer $K-K'$ is neglected and intra-
and inter-valley density-density interactions are set to be equal.
Because of the SU(4) symmetry of the model we may, without loss of generality, choose a basis which is diagonal in both the normal and ordered states, and describe the ordered states in terms of the densities of each of four bands $n_{\lambda\rho}$, where $\lambda$ and $\rho$ have plus or minus values, depending, respectively, on spin projection and valley number.
Similar to the two component case, we parameterize $n_{\lambda \rho}$ in terms of three variables
$\zeta_1, \zeta_2$ and $\zeta_3$, such that $n_{\lambda\rho} = n_{0}(1 + \lambda \zeta_1 + \rho \zeta_2 + \lambda \rho \zeta_3)$
where $n_{0}$ is the normal state density per spin and valley.
The total density is $n = 4 n_0$.
In these notations, the Landau
energy is
\begin{equation}
    F(\zeta_1,\zeta_2, \zeta_3)
    =
    \frac{1}{2}(\zeta_1^{2} +\zeta_{2}^{2} + \zeta_3^{2})
    \left(
    h(\zeta_{1},\zeta_{2},\zeta_{3})
    - \nu_{F,\alpha} U
    \right),
    \label{eq:du-su4}
\end{equation}
where
\begin{equation}
    h(\zeta_{1},\zeta_{2}, \zeta_{3})\equiv
    \frac{\sum_{\lambda\rho}(1 +\lambda \zeta_1 + \rho \zeta_2 + \lambda \rho \zeta_3)^{\alpha+1} - 4}{2\alpha(\alpha+1)(\zeta^{2}_{1}+\zeta^{2}_{2}+\zeta^{2}_{3})}.
    \label{eq:h-def-su4}
\end{equation}
As in the SU(2) case, upon reaching a threshold coupling, the first transition will be to the state that
minimizes $F$, subject to the constraints $0 \leq n_{\lambda\rho} \leq 4$.
We emphasize that this is the full
internal energy, valid for arbitrary $|\zeta_i| \leq 1$.
In the Taylor expansion,
$F(\zeta_1,\zeta_2, \zeta_3)$ has terms quadratic, cubic, quartic in $\zeta_i$, and so on~\cite{Chichinadze2022}
(cubic terms of the form $\zeta_1 \zeta_2 \zeta_3$ and higher-order odd-power terms are allowed by $\mathrm{SU(4)}$ symmetry)~\cite{Chichinadze2022}
\footnote{The authors of \cite{Chichinadze2022} expressed the Landau energy to quartic order in the order parameter in terms of three components of the traceless $\mathrm{SU(4)}$ matrix $\hat{\Phi}$ in the diagonal basis: $\hat{\Phi} = \operatorname{diag}(\lambda_1, \lambda_2,\lambda_{3}, -(\lambda_1+\lambda_2+\lambda_{3}))$.
The relation between $\zeta_i$, which we use in this paper, and $\lambda_i$ is $\zeta_1 = \lambda_2+\lambda_3, \quad\zeta_2 = \lambda_1+\lambda_3, \quad\zeta_3 = \lambda_1+\lambda_2$.}

\begin{figure}
    \centering
    \includegraphics[width=\linewidth]{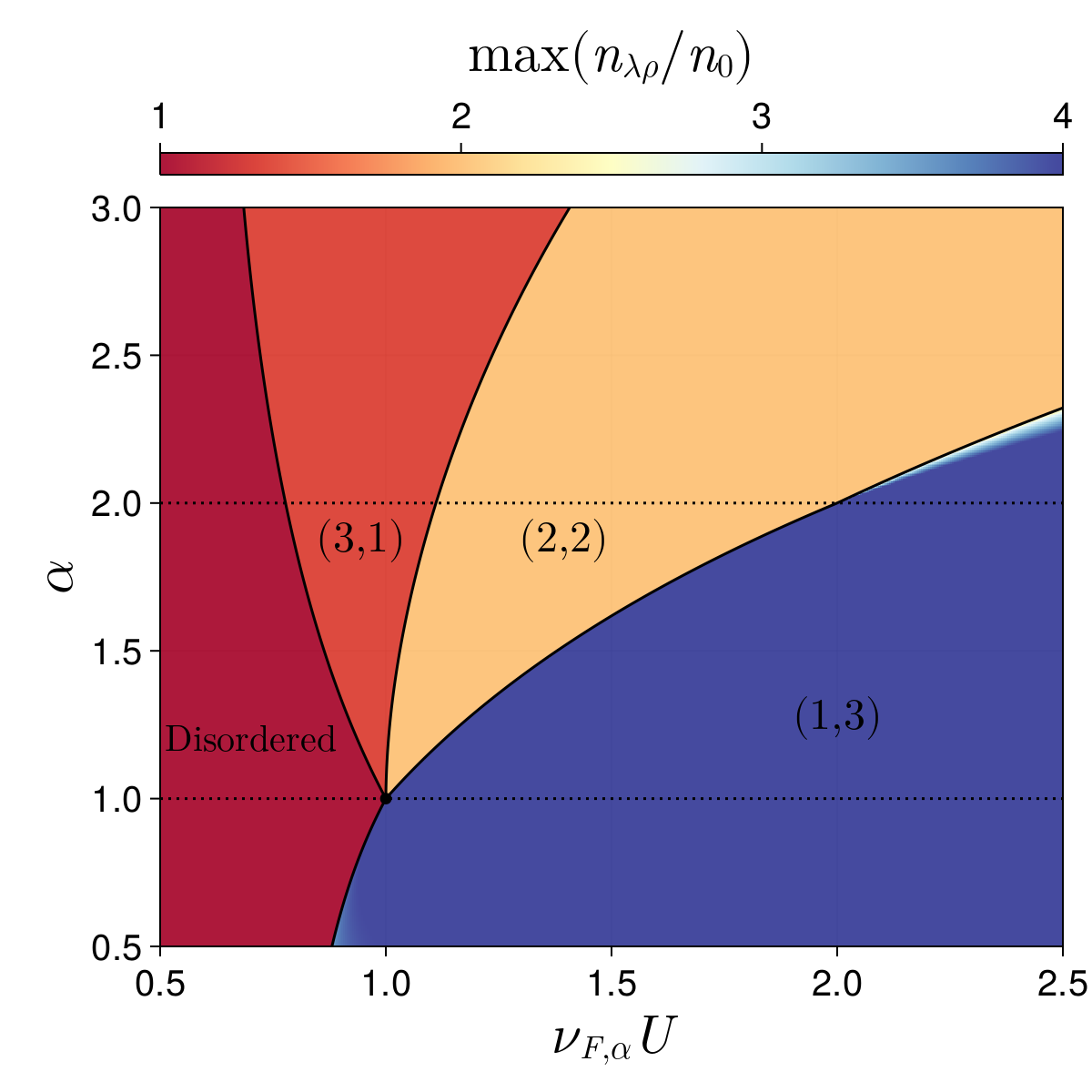}
    \caption{(Color online)  Phase diagram of the
        $\mathrm{SU(4)}$ invariant model. The polarization is incomplete in the lighter-colored regions with blurred boundaries.
        \label{fig:su4-phase}
    }
\end{figure}
\begin{figure}
    \centering
    \includegraphics[width=0.9\linewidth]{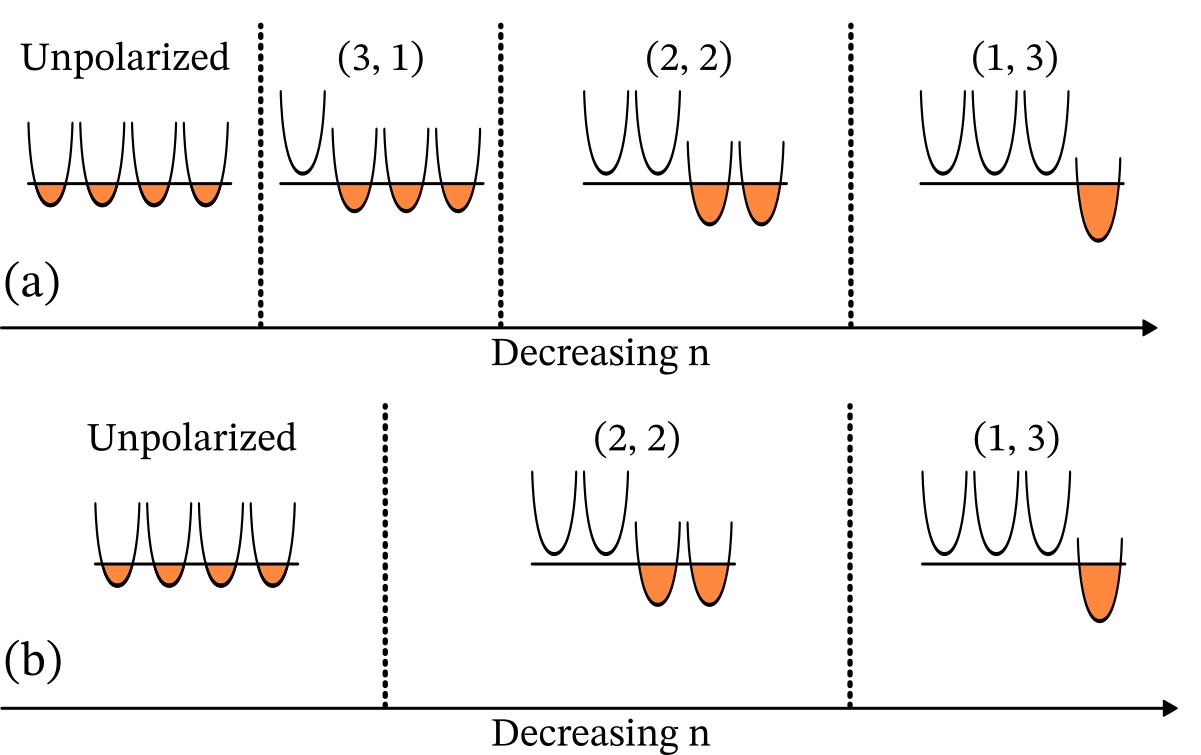}
    \caption{(Color online)
        The sequence of ordered states with decreasing density for $1 \leq \alpha \le 2$ in the (top) $\mathrm{SU(4)}$ invariant model and (bottom) anisotropic $\mathrm{SU(4)}$ model.\label{fig:su4_anisotropic} }
\end{figure}

The ground state of the system for any particular value of $\alpha$ is determined by the competition between the quadratic term and higher order even and odd terms.
We show the result of our analysis of $F(\zeta_1,\zeta_2, \zeta_3)$ in \cref{fig:su4-phase}.
For $1 < \alpha  < 2$, there is a cascade of transitions into a three-quarter-metal, a half-metal, and a quarter-metal states with
band densities $\{\frac{4}{3}n_{0},\frac{4}{3}n_{0}, \frac{4}{3}n_{0}, 0\}$, $\{2n_{0}, 2n_{0}, 0, 0\}$, and $\{4n_{0}, 0, 0, 0\}$, respectively (we label them as $(3,1)$, $(2,2)$ and $(1,3)$ states).
We show the corresponding band fillings in \cref{fig:su4_anisotropic}a.
These transitions occur at three progressively increasing interaction strengths, $U_{(3,1)} =  6\left([4/3]^{\alpha} - 1\right)/(\nu_{F,\alpha} \alpha[\alpha+1])$;  $U_{(2,2)} = 3\cdot 2^{\alpha}\left(1-[2/3]^{\alpha}\right]/ (\nu_{F,\alpha} \alpha[\alpha+1])$, and $ U_{(1,3)} = 2^{\alpha}(2^{\alpha}-1)/(\nu_{F,\alpha}\alpha[\alpha+1])$.
All transitions are first order, into maximal polarization, which gives rise to subsequent depletion of $1$, $2$, or $3$ bands out of 4.
Near each transition, the corresponding
susceptibility almost diverges, like for the ferromagnetic transition in the SU(2) case.
There is a true divergence for $\alpha =1$.
In terms of the 15 order parameters, the $(3,1)$ and $(1,3)$ states are necessarily mixed orders, e,g., valley polarization and ferromagnetism in just one valley~\cite{Chichinadze2022a}, while the order in the $(2,2)$  state can be a ``pure'' one, i.e., either valley polarization of intra-valley ferromagnetism, or inter valley coherence.
Collective excitations in the $(3,1)$ state are $8$ over-damped longitudinal modes of relative density fluctuations of the occupied Fermi surfaces, and $3$ Goldstone modes.
In the $(2,2)$ state, there are $3$ over-damped longitudinal modes, and $4$ Goldstone modes, and in the $(1,3)$ state there are just $3$ Goldstone modes. All the Goldstone modes are ``type-B''~\cite{Watanabe2011,*Watanabe2020}, \emph{i.e.},  have a $\omega\sim q^2$ dispersion, similar to the magnons in the SU(2) case.
In addition, in each state there is a plasmon mode in the total density channel.
The total number of modes in each state is less than in the normal state
as some modes disappear when one (or two or three) bands get fully depleted.

For $\alpha <1$, there is a single first-order transition at $U_{(1,3)}$ into a $(1,3)$  state.
For $\alpha < 0.89$, there is a very tiny range of $U<U_{(1,3)}$ where polarization is large but not maximal.
Similarly, for $\alpha >2$, the first order transitions into the $(3,1)$ and $(2,2)$ states persist at the same critical $U$ as for smaller $\alpha$, but the transition from the $(2,2)$ state to the $(1,3)$ becomes second order into the intermediate state with fillings
$\{2n_{0}+\delta n, 2n_{0}-\delta n, 0, 0\}$.
The range of intermediate polarization is, however, very narrow (see \cref{fig:su4-phase}).

\emph{Anisotropic $\mathrm{SU(4)}$ model.}~When $\mathrm{SU(4)}$ symmetry is broken, or when spin-orbit coupling is sizable~\cite{Xie2023,Szabo2022}, the range of the $(3,1)$ phase shrinks and eventually disappears, and the leading instability is into the half-metal $(2,2)$ state, like we found for a ferromagnetic Stoner transition, although in a spin/valley system the first instability can be into any of the possible ordered states.
Because
interactions in both valley and spin channels are attractive, above the leading instability there must be a secondary instability for the ``other'' isospin component (e.g., spin component, if the leading instability is in the valley sector).
This necessarily leads to a subsequent transition into the $(1,3)$ state.
We show the sequence of the ordered states in \cref{fig:su4_anisotropic}b.

\emph{Comparison with experiments.}~There are two parameters in our theory: the exponent $\alpha$ and the dimensionless coupling $U \nu_{F,\alpha}$.
The value of $\alpha$ is controlled by the displacement field $D$: in absence of trigonal warping fermionic dispersion can be approximated as $\epsilon \approx \sqrt{(k^{2}/2m)^{2} + D^{2}}$~\cite{McCann2013,Dong2023}, which interpolates between the $\alpha=2$ case at small $k$ and $\alpha=1$ at large $k$.
For these $\alpha$, the dimensionless coupling is a decreasing function of density, and our results predict that as density decreases
at a fixed $D$,
the $\mathrm{SU(4)}$-symmetric system undergoes a sequence of first order transitions.
In the SU(4) symmetric case, the state evolves
from a spin and valley symmetric $(4,0)$ metallic state at larger densities to $(3,1)$, $(2,2)$ and $(1,3)$ states.
In the presence of anisotropy,
the
range of the $(3,1)$ state shrinks, and for large enough anisotropy
the
sequence of first order transitions is $(4,0) \to
(2,2) \to
(1,3)$,
see~\cref{fig:su4_anisotropic}.
This is consistent with what has been observed experimentally in BBG and RTG~\cite{Zhou2021,Seiler2022,Zhou2022a,DeLaBarrera2022,Seiler2023,Arp2023,Holleis2023,Zhang2023a}.
We also found, for $\alpha >2$ and $\alpha <1$,
narrow ranges of intermediate states with depleted but not empty bands.
Such intermediate states have also been observed in experiments~\cite{Zhou2021,Zhou2022a,DeLaBarrera2022, Arp2023,Holleis2023,Zhang2023a}.
Our results also shed some light on the interplay between metallic states at fractional electronic fillings and Chern-insulating states near integer fillings.
Namely, metallic behavior at some generic non-integer filling likely can be described by confining to low-energy states near the chemical potential and neglecting fermionic excitations on the other side of the Dirac point.
If order developed continuously, it would then be difficult to describe how it can lead to a Chern insulator.
However, our results show that the magnitude of the spin/valley order parameter instantly jumps to its maximal possible value.
Such an order affects fermions at all energies, including states on the other side of the Dirac point.
Possibly, this is a prerequisite for Chern-insulating behavior.

\emph{Conclusions.}~We have demonstrated that the ferromagnetic Stoner transition for 2D itinerant fermions is highly unconventional
for a range of 2D models with power-law fermionic dispersions $k^{2\alpha}$.
Namely, the order parameter susceptibility diverges, or almost diverges at the onset of order, as expected for a continuous transition, yet immediately beyond the instability an order parameter jumps to its maximal value, completely depleting one of the spin bands.
We extended the analysis to models with spin and valley degrees of freedom and argued that for a range of dispersions the ordered states are half-metal and quarter metal (and possibly three-quarter metal) with $1, 2$, or $3$ bands out of $4$ fully depleted.
We argue that these quantized states come about because of the form of the functional dependence of the density of states on energy.
The found sequence of transitions is consistent with the one
found in the experiments on BBG and RTG.\@
The sequence of transitions into spin/valley ordered states
in BBG and RTG
has been obtained in earlier
theoretical works~\cite{Zondiner2020,Zhou2021,Chichinadze2022,Chichinadze2022a,Rakhmanov2023,*Rakhmanov2023a,*Rakhmanov2023b}.
The novel aspect of our results is the derivation and analysis of the full Landau internal energy for competing spin and valley order parameters, without assuming that order parameter magnitudes are small
or selecting \emph{a priori} a particular spin/valley order.
We also note that similar discontinuous transitions  into half-metal and quarter-metal states have been recently detected in a 2D electron gas $AlAs$~\cite{Hossain2020,*Hossain2021,*Hossain2022,*Valenti2023}.

\begin{acknowledgments}
We thank E. Berg, Z. Dong, C. Chamon, D. Chichinadze, L. Classen, I. Esterlis, F. Guinea, S-S Lee, L. Levitov, H. Ma, A.MacDonald, S. Nadj-Perge, Y. Oreg, A. Paramekanti, A-M Tremblay, Y. Wang,  T. Weitz, J. Wilson and A. Young for fruitful discussions and feedback.
The work of AVC was supported by U.S.
Department of Energy, Office of Science, Basic Energy Sciences, under Award No. DE-SC0014402;
LIG acknowledges the support by NSF Grant No. DMR-2410182 and by the Office of Naval Research (ONR) under Award No. N00014-22-1-2764.
Part of the work was performed while the authors visited the Kavli Institute for Theoretical Physics (KITP).
KITP is supported in part by grant NSF PHY-1748958.
\end{acknowledgments}

\bibliography{references.gen}
\end{document}